\documentclass[paper=letter, fontsize=12pt]{article}

\usepackage[english]{babel} 
\usepackage{amsmath,amsfonts,amsthm} 
\usepackage[utf8]{inputenc}
\usepackage{float}
\usepackage{lipsum} 
\usepackage{blindtext}
\usepackage{graphicx} 
\usepackage{caption}
\usepackage[sc]{mathpazo} 
\usepackage[T1]{fontenc} 
\usepackage{microtype} 
\usepackage[hmarginratio=1:1,top=32mm,columnsep=20pt]{geometry} 
\usepackage{multicol} 
\usepackage{booktabs} 
\usepackage{float} 
\usepackage{hyperref} 
\usepackage{lettrine} 
\usepackage{paralist} 
\usepackage{abstract} 
\usepackage{titlesec} 
\usepackage{natbib, url}
\usepackage[caption=false]{subfig}   
\usepackage{algorithm} 
\usepackage{algpseudocode}

\renewcommand\thesection{\Roman{section}} 
\renewcommand\thesubsection{\Roman{subsection}} 

\titleformat{\section}[block]{\large\scshape\centering}{\thesection.}{1em}{} 
\titleformat{\subsection}[block]{\large}{\thesubsection.}{1em}{} 
\usepackage{fancyhdr} 
\usepackage[bottom]{footmisc}

\newtheorem{theorem}{Theorem}[section]

\numberwithin{equation}{section}

\renewcommand\thesection{\arabic{section}}
\renewcommand\thesubsection{\thesection.\arabic{subsection}}

\graphicspath{{./pics/}}

\title{\vspace{-15mm}\fontsize{24pt}{10pt}\selectfont\textbf{Applications of Conjugate Gradient in Bayesian computation}} 

\author{
\large
{\textsc{Lu Zhang}}\\[2mm]
{\textsc{Division of Biostatistics}}\\[2mm]
{\textsc{Department of Population and Public Health Sciences}}\\[2mm]
{\textsc{University of Southern California}}\\[2mm]
\normalsize \href{mailto:lzhang63@usc.edu}{lzhang63@usc.edu}\\[2mm] 
}
\providecommand{\keywords}[1]{\textbf{\textit{Key words:}} #1}
\begin{document}
\maketitle 
\thispagestyle{fancy} 

\label{firstpage}

\begin{abstract}
Conjugate gradient is an efficient algorithm for solving large sparse linear systems. It has been utilized to accelerate the computation in Bayesian analysis for many large-scale problems. This article discusses the applications of conjugate gradient in Bayesian computation, with a focus on sparse regression and spatial analysis. A self-contained introduction of conjugate gradient is provided to facilitate potential applications in a broader range of problems. (This paper was originally published on Wiley StatsRef: Statistics Reference Online on December 15 2022. The reason for reuploading it on arXiv is to enhance its visibility and accessibility.)
\end{abstract}

\keywords{conjugate gradient; preconditioning; Bayesian computation; sparse regression; spatial analysis}

\section{Introduction}\label{sec1}

Conjugate gradient (CG) \citep{hestenes1952methods, lanczos1952solution} is an iterative method for solving a linear system. It belongs to a broad class of iterative methods known as Krylov subspace methods.  Suppose that the target linear system is $Ax = b$ with $A$ an $n \times n$ matrix, and the corresponding solution is $x^\ast = A^{-1}b$ when $A$ is non-singular.
Rather than requiring matrix-matrix operations like elimination and direct decomposition algorithms, Krylov subspace methods access matrices only through matrix-vector multiplies and work with the resulting vectors. Therefore, the CG method is usually more efficient for solving large sparse system in that 1) it can avoid storing matrix factors or even $A$ and 2) it can utilize the sparsity of $A$ to reduce the storage and computational burden. Moreover, the CG method can be significantly faster than other direct solvers when $A$ has certain features, which will be discussed in Section~\ref{subsec: precond_CG}.  Readers can find comprehensive summaries of Krylov subspace methods in books like \citet{golub2012, datta2010numerical, hager2022applied, saad2003iterative, wright1999numerical}.

The CG method has been applied in many studies to facilitate computation in Bayesian analysis. Broadly speaking, the CG method, as well as other iterative methods, serves two functions in Bayesian computation, evaluation of an expensive function and sampling of a large dimensional distribution. 
Typical applications can be found in research about sparse regression and large scale spatial data analysis. CG has been utilized to evaluate the Bayes factors and generate cheap updates in Markov chain Monte Carlo (MCMC) sampling for sparse regression models. CG has also found applications in the evaluation of hyper-parameters and the generation of posterior samples for Gaussian process models in spatial analysis. A more detailed summary and discussion about these applications is presented in Section~\ref{sec: app}. 
In Section~\ref{sec: CG}, an introduction of the CG method and a summary of some of the most practically useful results about CG are provided in a concise and self-contained manner.

\section{Introduction of conjugate gradient methods}\label{sec: CG}
\subsection{Conjugate gradient}\label{subsec: CG}
Consider the problem of solving the large linear system $Ax = b$ where $A$ is an $n \times n$ matrix. Assume $A$ is non-singular, and denote its inverse as $A^{-1}$. From a theoretical point of view, we can further assume $A$ matrix to be symmetric and positive definite, since the solution of system $Ax = b$ is equal to that of $Bx = t$ when $B = A^\top A$, $t = A^\top b$ with $A^\top$ being the transpose of $A$.  

The CG method transforms the problem of solving the linear system $Ax = b$ into one of minimizing a quadratic function $\phi(x) = \frac{1}{2}x^\top A x - x^\top b$. With an initial guess $x_0$, CG generates a sequence of approximate solutions $x_{0}, x_{1}, \ldots$ that converges to $x^\ast = A^{-1}b$ in at most $n$ iterations. Specifically, starting from an initial search direction $p_0 = r_0 = b - A x_0$, we update $x_{k-1}$ for $k = 1, 2, \ldots$ by searching along the direction $p_{k-1}$ to find $x_{k} = x_{k-1} + a_{k-1} p_{k-1}$ that minimizes the target quadratic function $\phi(x)$ on the search line. The optimal step size $a_{k-1}$ in the update turns out to be $p_{k-1}^\top r_{k-1}/p_{k-1}^\top A p_{k-1}$, and the residual at iteration $k$ ($r_{k} = b - Ax_{k}$) can be calculated by $r_{k} = r_{k -1} - a_{k-1} A p_{k-1}$. The search direction for the next iteration $p_{k}$ is generated based on $p_{k-1}$ by the formula
 \begin{equation}\label{eq: CG_p_update}
 p_{k} = r_{k} + \tau_{k-1} p_{k-1}\;, \; \tau_{k-1} = - r_{k}^\top A p_{k-1}/ p_{k-1} A p_{k-1}\;.
 \end{equation}
 It can be proved that the searching directions $p_0, p_1, \ldots$ are mutually conjugate, and the residuals $r_0, r_1, \ldots$  are mutually orthogonal, that is, 
 $$
  p_i^\top A p_j = 0\;, \; r_i^\top r_j = 0 \;, \; (i \neq j)\;
 $$
 \citep[see Section~5 in ][]{hestenes1952methods}. 
 The CG method is designed so that $x_{k}$ minimizes the target quadratic function $\phi(x)$ over a space spanned by $\{r_{0}, \ldots, r_{k-1}\}$. Note that the residual $r_{k-1} = b - A x_{k-1}$, which equals the negative gradient of $\phi(x)$ at $x_{k-1}$, is the direction where the target function $\phi(x)$ decreases most rapidly at $x_{k-1}$; this guarantees that a CG update is at least as good as a steepest descent update. 
 
 Algorithm~\ref{alg1} delineates an economic version of CG. The procedure is essentially the CG algorithm form described in Section~11.3.8 of \citet{golub2012}, with modifications to better trace the storage and computational burden. Each CG step in Algorithm~\ref{alg1} requires updating four n-vectors, and the matrix $A$ is involved only in one matrix-vector product. These features make CG applicable for solving large sparse linear systems.

\begin{algorithm}
\caption{Conjugate gradient method}\label{alg1}
\begin{algorithmic}
\State Initialize $k = 0$, $x = x_0$, $p = r = b - Ax$, $\rho_c = r^\top r = \rho_{-}$, $\tau = 1 = a$. 
\State Pre-allocate the $n$-vector $w$
 \While{convergence criteria is unmet, e.g. $\rho_c > 0$}
 \State $k = k + 1$
 \If{$k > 1$}{ $\tau = \rho_c / \rho_{-}$,  $p = r + \tau p$ \hfill{$\mathcal{O}(n)$ flops, one saxpy}}
 \EndIf
 \State $w = Ap$ (only step involves matrix $A$)
 \State $a = \rho_c / p^\top w$, $x = x + a p$, $r = r-aw$, $\rho_{-} = \rho_c$, $\rho_c = r^\top r$ \hfill{$\mathcal{O}(n)$ flops, two saxpys and two inner products}
 \EndWhile
 \State {(Note: ``saxpy'' stands for ``constant times a vector plus a vector'', i.e., the routine that updates $y$ by $y = ax + y$)}
\end{algorithmic}
\end{algorithm}

\subsection{Convergence rate and preconditioned CG}\label{subsec: precond_CG}
We have seen the properties of CG that are beneficial for solving large sparse systems. More remarkably, the CG algorithm can identify the solution in many fewer than $n$ iterations when the distribution of the eigenvalues of $A$ has certain favorable features. For example, the following theorem summarizes some important findings in \citet[p299][]{trefethen1997numerical} 
and 
\cite{luenberger1973introduction}. These conclusions about error bounds give a useful characterization of the behavior of the CG method. 
\begin{theorem}\label{thm1}
Let $0 < \lambda_1 \leq \lambda_2 \leq \ldots \leq \lambda_n$ be the ordered eigenvalues of the positive definite matrix $A$. Defining the norm measure $\|\cdot\|_A$ by $\|u\|_A^2 = u^\top A u$, we have that
\begin{equation}\label{eq: thm1_1}
\|x_{k+1} - x^\ast\|_{A}^2 \leq \frac{\lambda_{n - k} - \lambda_1}{\lambda_{n-k} + \lambda_1} \|x_0 - x^\ast\|_{A}^2\;, \mbox{ and }
\end{equation}
\begin{equation}\label{eq: thm1_2}
\|x_{k} - x^\ast\|_{A} \leq 2\|x_{k} - x_{0}\|_A \left(\frac{\sqrt{\kappa(A)} - 1}{\sqrt{\kappa(A)} + 1}\right)^k\;.
\end{equation}
where $\kappa(A) = \lambda_n / \lambda_1$ is the conditional number of matrix $A$.
\end{theorem} 
More intuitively, Theorem~\ref{thm1} implies that 1) if the eigenvalues of $A$ occur in $h$ distinct clusters, then the solution produced by the CG method becomes very close to $x^\ast$ in $O(h)$ iterations \citep[see discussions on p117 in][]{wright1999numerical}, and 2) early termination can be expected when the condition number $\kappa(A)$ is small. 

An important strategy for accelerating the CG method is preconditioning, which can improve the eigenvalue distribution of $A$. Preconditioning works by modifying CG to solve the related linear system 
$
(C^{-\top} AC^{-1}) C x = C^{-\top} b,
$
instead of solving the original system $Ax = b$ (where $C^{-\top}$ is the inverse of the transpose of matrix $C$). The modification does not change the solution of the linear system. However, the convergence rate of CG depends on the eigenvalue distribution of $C^{-\top} AC^{-1}$ rather than that of $A$. The implementation of the preconditioned CG method does not utilize $C$ explicitly. It makes use of the preconditioner $M = C^\top C$ as described in Algorithm~\ref{alg2}. The boxed equations in Algorithm~\ref{alg2} highlight the differences between the preconditioned and unpreconditioned CG methods. 

\begin{algorithm}
\caption{Preconditioned conjugate gradient method}\label{alg2}
\begin{algorithmic}
\State Initialize $k = 0$, $x = x_0$, $r = b - Ax$, $\boxed{p = z = M^{-1} r}$, $\boxed{\rho_c = r^\top z = \rho_{-}}$, $\tau = 1 = a$. 
\State Pre-allocate the $n$-vector $w$
 \While{convergence criteria is unmet, e.g. $\|r\|_2 > 0$}
 \State $k = k + 1$
 \If{$k > 1$}{ $\tau = \rho_c / \rho_{-}$,  $\boxed{p = z + \tau p}$ }
 \EndIf
 \State $w = Ap$ 
 \State $a = \rho_c / p^\top w$, $x = x + a p$, $r = r-aw$, 
 $\rho_{-} = \rho_c$, $\boxed{z = M^{-1}r}$, $\boxed{\rho_c = r^\top z}$
 \EndWhile
\end{algorithmic}
\end{algorithm}

A good preconditioner $M$ should be selected to improve the eigenvalue distribution of $C^{-\top} A C^{-1}$ as well as be inexpensive to compute $z = M^{-1}r$. Several general-purpose preconditioners have been proposed. Typical strategies include symmetric successive overrelaxation (SSOR), incomplete Cholesky, sparse approximate and inverse preconditioners. For more discussion about those techniques, we refer readers to \citet{golub2012, wright1999numerical, saad2003iterative}. 
There is no single preconditioning strategy that is ``best'' for all problems. We will see in the next section that an efficient implementation of CG in Bayesian computation often requires a dovetailed preconditioner. 

\section{Applications of CG in Bayesian computation}\label{sec: app}
This section is devoted to a summary of representative applications of CG in Bayesian computation. The presentation mainly focuses on the applications in sparse regression and spatial analysis. Brief background introductions are given, followed by a summary of CG applications.  

\subsection{Sparse regression}
Assume that we have $n$ data points, each records the value of outcome $y_i$ and the corresponding values of the $p$ covariates $x_{i1}, x_{i2}, \ldots, x_{ip}, i = 1, \ldots, n$. A typical linear regression model is
$$
y = X \beta + \epsilon\;, \; \epsilon \sim \mbox{MVN}(0, \tau^2 I_n)\;,
$$
where $y = (y_1, \ldots, y_n)$ is the vector of outcomes, $X = \{x_{ij}\}_{i = 1, \ldots, n, j = 1, \ldots, p}$ is the $n \times p$ design matrix, $\epsilon = (\epsilon_1, \ldots, \epsilon_n)$ is an $n$-vector whose elements $\epsilon_i \sim \mbox{N}(0, \tau^2)$ for $i = 1, \ldots, n$ and $I_n$ is an identity matrix with dimension $n$. When both the sample size $n$ and the number of covariates $p$ are large, sparse regression can be implemented to find a small subset of covariates that captures the main relationship between the outcome and the covariates. The sparse regression has been used in various applications such as modern observational studies for healthcare related databases and genome-wide genetic data analyses. Sparse regression is also known as penalized regression and shrinkage regression.
In a Bayesian paradigm, the core modeling strategy of sparse regression is to assign a structured prior for the high-dimensional regression coefficient $\beta$ so that the number of no-zero elements in the posterior samples of $\beta$ can be controlled to be small.  The ``spike-and-slab'' priors and shrinkage priors are popular options. Obtaining the posterior inference often requires an MCMC sampling process. The MCMC sampling can be extremely expensive for such a ``large $p$ \& large $n$'' problem, hampering the performance of Bayesian sparse regression. 

A fruitful direction for improving the model fitting of Bayesian sparse regression is to exploit the sparsity structures in the priors and, if any, the special structures of the design matrix $X$ to obtain a more efficient sampling process. A representative example is \citet{zhou2019fast}. \citet{zhou2019fast} proposed the iterative complex factorization algorithm for solving a special class of large linear systems and utilized it to ease the computation of Bayes factors required in an MCMC algorithm for a large linear regression with a ``spike-and-slab'' prior for $\beta$. In addition to facilitating the evaluations required by MCMC, iterative methods can also be used to accelerate MCMC by efficiently producing samples. 
\citet{nishimura2022prior} devised a Gibbs sampler for a sparse logistic regression model with a shrinkage prior for regression coefficients. They recast the problem of sampling from the conditional posterior distribution of regression coefficients as that of solving a linear system, and made use of CG to speed up the posterior sampling. The authors developed the prior-preconditioning strategy and provided theoretical results to show that the CG method with their prior-preconditioner is superior over general-purpose preconditioners in Bayesian sparse regression applications.



\subsection{Spatial analysis}
Rapidly increasing usage and growing capabilities of Geographic Information Systems (GIS) have spawned considerable
research in modeling and analyzing large scale spatial data sets \citep[see application examples in][]{gelfand2010handbook, cressie2015statistics, banerjee2003hierarchical}. Gaussian
process (GP) is a widely used modeling tool in spatial analysis, which offers flexibility and richness in modeling a spatial random field \citep{stein1999interpolation}. However, fitting GP models incurs onerous computational costs that severely hinders the implementation of GP models for large data. The key bottleneck stems from the massive spatial covariance matrix to model the spatial correlation among observations. For irregularly situated spatial locations, as is common in geostatistics, the covariance matrix is typically dense and carry no exploitable structure to facilitate computations.
Since the storage and computational costs of Cholesky decomposition required in the likelihood evaluation are in the order of square and cubic of the number of locations, respectively, the implementation quickly become infeasible as the number of locations increase. 
Even for a modestly large number of points ($\approx 50,000$ or greater), the computational demands become prohibitive for a modern computer and preclude inference from GP models.

The expensive linear algebra operations, such as multiplying two matrices or factorizing a matrix, in optimizing the hyper-parameters of a GP model can be avoided by solving linear systems involving large covariance matrices \citep{gibbs1997efficient}. Based on that, the iterative methods like CG have been used in some significant works under the focus of parameter estimation.
For example, \citet{stein2012difference} explored the preconditioning techniques based on a differencing approach to precondition the covariance matrix in score equations (i.e. score function equal to $0$), and implemented the iterative algorithm to compute the hyper-parameter estimates. 
\citet{sun2016statistically} utilized preconditioned CG to find the maximum likelihood estimator of hyper-parameters by solving estimating equations based on score equation approximates. 
CG has also been used in the simulation of the exact GP models.
\cite{wang2019exact} leveraged multi-GPU parallelization and methods like CG to train an exact GP on over a million points. Their implementations use the partial pivoted Cholesky decomposition introduced by \cite{gardner2018gpytorch} as a preconditioner for the covariance matrix of response process observed with white noise. 

The applications of CG in Bayesian spatial analysis mainly contribute to the posterior sampling procedure. 
\citet{stroud2017bayesian} proposed a two-block Gibbs sampler that alternates between updating the missing data and the parameters for GP on an incomplete lattice. A preconditioned CG method was implemented to update the posterior samples of the missing data in the proposed Gibbs sampler. The authors experimented a number of preconditioners to spot the best preconditioner built based on Vecchia's approximation \citep{vecchia1988estimation}.
\citet{zhang2021spatial} constructed scalable versions of spatial factor models for large-scale multivariate spatial data analysis through a combination of linear model of coregionalization and a number of univariate scalable spatial modeling strategies. The proposed models deliver inference over a high-dimensional parameter space, including the latent spatial processes on observed and unobserved locations, via a block update MCMC algorithm. CG is used to generate updates for the latent spatial processes in the MCMC algorithm. Similar applications can be found in \citet{zhang2019practical, zhang2021high} where the authors proposed pragmatic spatial analyzing methods by fitting scalable conjugate spatial models with cheap posterior sampling through CG.

\section{Conclusions}\label{sec: dicussion}
The CG method provides a useful tool to solve large sparse linear systems. There are many applications in fields, such as sparse regression and spatial analysis, from which we have seen that the usage of CG in Bayesian computation can be quite flexible. For specific problems, an efficient CG application needs to design the implementation, exploit the sparsity in the model, and tailor the corresponding preconditioning strategy. CG has the potential to find more applications in a broader range of statistical problems.

\section*{Acknowledgements}
This work is supported by the U.S. National Science Foundation (grant 2055251) and the U.S. Office of Naval Research. I especially thank Akihiko Nishimura, who recommended me to the editor for writing this article.

\appendix

\bibliographystyle{ba}  
\bibliography{WileySTAT} 

\end{document}